\documentclass[showpacs,preprintnumbers,amsmath,amssymb,prb]{revtex4}

\def\XXint#1#2#3{{\setbox0=\hbox{$#1{#2#3}{\int}$}
     \vcenter{\hbox{$#2#3$}}\kern-.5\wd0}}

\usepackage{inputenc}
\usepackage{graphicx}
\usepackage{subfigure}
\usepackage{bm}

\begin{document}

\title{Magnetically ordered phase near transition to Bose-glass phase}

\author{A.\ V.\ Syromyatnikov$^{1,2}$}
\email{asyromyatnikov@yandex.ru}
\author{A.\ V.\ Sizanov$^{1}$}
\email{alexey.sizanov@gmail.com}
\affiliation{$^1$National Research Center "Kurchatov Institute" B.P.\ Konstantinov Petersburg Nuclear Physics Institute, Gatchina 188300, Russia}
\affiliation{$^2$Saint Petersburg State University, 7/9 Universitetskaya nab., St. Petersburg, 199034 Russia}

\date{\today}

\begin{abstract}

We discuss magnetically ordered (``superfluid'') phase near quantum transition to Bose-glass phase in a simple modeling system, Heisenberg antiferromagnet in spatial dimension $d>2$ in external magnetic field with disorder in exchange coupling constants. Our analytical consideration is based on hydrodynamic description of long-wavelength excitations. Results obtained are valid in the entire critical region near the quantum critical point (QCP) allowing to describe a possible crossover from one critical behavior to another. We demonstrate that the system behaves in full agreement with predictions by Fisher et al.\ (Phys.\ Rev.\ B {\bf 40}, 546 (1989)) in close vicinity of QCP. We find as an extension to that analysis that the anomalous dimension $\eta=2-d$ and $\beta=\nu d/2$, where $\beta$ and $\nu$ are critical exponents of the order parameter and the correlation length, respectively. The density of states per spin of low-energy localized excitations are found to be independent of $d$ (``superuniversal''). We show that many recent experimental and numerical results obtained in various 3D systems can be described by our formulas using percolation critical exponents. Then, it is a possibility that a percolation critical regime arises in the ordered phase in some 3D systems not very close to QCP.

\end{abstract}


\maketitle

\section{Introduction}

The problem of interacting bosons in disordered potential (the so-called dirty-boson problem) has been an active field in recent years beginning with the seminal paper by Fisher et al. \cite{Fisher} The recent upsurge of interest to dirty bosons is stimulated by their novel experimental realizations in ultracold atoms in optical lattices with controlled disorder (see, e.g., Ref.~\cite{Sanchez}) and in doped quantum magnets (see Ref.~\cite{Zhel13} for review).

The most pronounced effect of disorder is the localization of single-particle low-energy states which cannot host a Bose-Einstein condensate at finite interaction between bosons. As a result, a novel non-superfluid glassy phase arises, the so-called Bose-glass (BG) phase, which intervenes between the superfluid (SF) and Mott insulating (MI) ones and which destroys MI phase at strong enough disorder. \cite{Fisher} Due to finite density of states of localized gapless quasiparticles, BG phase (as SF one) has a finite compressibility. In contrast, MI phase is incompressible. The transition from MI to BG phases is found to be of the Griffiths type. \cite{Fisher,grif1,grif} Properties of the transition from SF to BG phases are in hot debate now. The situation is greatly complicated by the fact that many conventional approaches are not appropriate for dirty bosons. In particular, a perturbative renormalization group treatment breaks down because the upper critical dimension is infinitely large. \cite{Fisher} As a consequence, a rigorously motivated scaling form is absent of the free energy so that the simplest scaling ansatz proposed in Ref.~\cite{Fisher} 
is called in question recently \cite{yu2,weich} in view of some apparent contradictions to numerical and experimental data. 

It is shown in Ref.~\cite{Fisher} that the finiteness of the compressibility at the quantum critical point (QCP) between SF and BG phases implies that the dynamical critical exponent $z$ is equal to the spatial dimension of the system $d$. Numerical results obtained in the largest 2D \cite{2d7} and 3D \cite{kisel} systems do support this conclusion. Besides, an alternative derivation of this relation is proposed in Ref.~\cite{yu2} which relies only on the assumption that the system is not above the upper critical dimension. Then, the equality $z=d$ seems to be well established at QCP.

Another consequence of the scaling ansatz proposed in Ref.~\cite{Fisher}, $\phi=\nu d$, is debated now, where $\phi$ and $\nu$ are critical exponents of the critical temperature $T_c\propto (x_c-x)^\phi$ and the correlation length $\xi\propto (x_c-x)^{-\nu}$ (here $x$ is a control parameter and $x_c$ is its value at QCP). There is a large amount of experimental and numerical data obtained in various 3D systems which are in sharp contradiction with this relation being quite consistent with each other (see Refs.~\cite{Zhel13,Yu,yu2,Huv1,wulf,yama,haas,critkis} and references therein). To resolve this discrepancy, a numerical investigation has been carried out recently in all-time large 3D systems \cite{kisel} which results are fully consistent with Ref.~\cite{Fisher}. It is claimed in Ref.~\cite{kisel} that previous studies were performed away from quantum critical region (in particular, because of too small system sizes considered in previous numerical works). This statement is disputed in Ref.~\cite{critkis}, where, in particular, the consistency is stressed between data obtained experimentally and numerically in various systems that testifies a universal behavior differing from that predicted in Ref.~\cite{Fisher}. As a possible source of discrepancy authors of Ref.~\cite{critkis} point out that the results of Ref.~\cite{kisel} characterize the neighborhood of one special point of the phase diagram in the plane of disorder strength and chemical potential whereas previous considerations which contradict Ref.~\cite{kisel} relate to another part of the phase diagram. Our results below could support a compromise idea: the existence of a crossover in a part of the phase diagram from the critical behavior proposed in Ref.~\cite{Fisher} to another one which is described by percolation critical exponents (or by critical exponents which are close to percolation ones). 

We discuss QCP between magnetically ordered (``SF'') and BG phases in a simple modeling system from the considered universality class, Heisenberg antiferromagnet (HAF) on hypercubic $d$-dimensional lattice whose Hamiltonian has the form
\begin{equation}
\label{haf}
 {\cal H} = \sum_{\langle i,j\rangle} J_{ij}\mathbf{S}_i \mathbf{S}_j - h\sum_i S^z_i,
\end{equation}
where $\langle i,j\rangle$ denote nearest-neighbor sites, $h$ is the value of the external magnetic field, and disorder in exchange coupling constants $J_{ij}>0$ is implied. As it is well known, model \eqref{haf} can be mapped onto an extended Bose-Hubbard model that is frequently used for discussion of dirty bosons (see, e.g., Ref.~\cite{Zhel13}). We develop a hydrodynamic description of long-wavelength excitations in the ordered phase whose spectrum is determined by static observable quantities: the static susceptibility and the helicity modulus, which is characterized by critical exponent $\sigma$. We describe model \eqref{haf} near QCP using general properties of a system in a critical regime and not involving any scaling ansatz. We demonstrate that model \eqref{haf} behaves in full agreement with predictions by Fisher et al.\ (Ref.~\cite{Fisher}) in the close vicinity of QCP. In particular, we recover many relations between critical exponents derived in Ref.~\cite{Fisher} from the scaling ansatz (including $\phi=\nu d$). As an extension to that analysis, we find that the anomalous dimension $\eta=2-d$ and the critical exponent of the order parameter $\beta=\nu d/2$ (the latter is in agreement with numerical results \cite{kisel}). The density of states (DOS) is also derived. DOS per spin of low-energy localized excitations are found to be ``superuniversal'' (i.e., independent of $d$). An alternative derivation is suggested of the equality $z=d$ at QCP based on common scaling arguments. We demonstrate that many recent experimental and numerical data obtained in various 3D systems can be described by our formulas using percolation critical exponents. This finding implies that a percolation critical regime can arise in the ordered phase not very close to QCP at $d=3$ which is characterized by critical exponents $z=1+\sigma/2\nu\ne d$ and $\phi=\sigma+(2-d)\nu\ne\nu d$. 

The rest of the present paper is organized as follows. We present a qualitative consideration of the problem and summarize our basic assumptions in Sec.~\ref{prel}. We remind briefly in Sec.~\ref{dilhaf} the approach applied before to diluted HAFs near the percolation threshold and adopt it for consideration of our system in Sec.~\ref{genqcp}. Formulas derived in Sec.~\ref{genqcp} which are valid in the whole critical region are applied in Sec.~\ref{qcp} for discussion of the close vicinity of QCP. The possible percolation scenario in some 3D systems is discussed in Sec.~\ref{seccross}. Sec.~\ref{conc} contains a summary and our conclusion. One appendix is included with some details of the analysis.

\section{Antiferromagnets with disorder near QCP}
\label{afdis}

\subsection{Qualitative consideration and basic assumptions}
\label{prel}

In the absence of disorder, a transition takes place in model \eqref{haf} at some field value $h=h_{c0}$ from magnetically ordered (``SF'') phase with Goldstone excitations to fully saturated (magnetically disordered ``MI'') phase with gapped spectrum (the gap value is given by $h-h_{c0}$). Mean transverse spin component plays the role of the order parameter in this transition. This quantum phase transition belongs to the universality class of Bose-Einstein condensation characterized by mean-field critical exponents $\phi=2/d$, $\beta=\nu=1/2$, and $z=2$. \cite{sachdev} 

We assume below that values of a small part of randomly chosen exchange constants $J_{ij}$ are increased. As a result, one has to apply a field greater than $h_{c0}$ to saturate spins around defect bonds. It is natural to expect also that local values of the saturation field are greater in regions with more dense distribution of defects. Then, the following qualitative picture arises. At $h<h_{c0}$, all spins have finite mean transverse components. As the field exceeds $h_{c0}$, less dense regions having smaller local critical fields become magnetically disordered (i.e., saturate) and the magnetically ordered part of the system acquires the form of an infinite network. Some regions (which are surrounded by areas with lower local critical fields) leave the infinite network before $h$ approaches their local critical field values. These regions, being isolated from each other and from the infinite network, do not contribute to the net order parameter of the sample which is determined solely by that of the infinite network. The number of sites decreases in the infinite network upon the field increasing and it disappears (falls to clusters of finite volume) at a critical field $h_c>h_{c0}$. This is the transition point from the magnetically ordered (``SF'') phase to the disordered BG one. It resembles qualitatively the conventional percolation transition that was pointed out before in many papers (see, e.g., Ref.~\cite{haas}). 

It is worth to remind that it is the self-similar (fractal) geometry of a lattice at the percolation threshold that makes the percolation transition to be universal. Then, geometric properties of a system are characterized by a unique characteristic length scale $\xi$ (the correlation length) which diverges at the percolation threshold. Many critical indexes of systems near percolation threshold are related to (universal) geometric characteristics of the random fractal (such as the fractal dimension, for example). We assume below that the universality of the transition from SF to BG phases has the same origin: finite clusters form a random fractal at QCP with some characteristics which differ in general from those of the percolation fractal. Geometric properties of regions with unsaturated magnetization are characterized close to QCP by the correlation length $\xi\propto (h_c-h)^{-\nu}$. We assume also (as in percolation theory) that one can find finite clusters of all characteristic linear sizes smaller than $\xi$ near QCP while the probability to find larger clusters is exponentially small.

Under such assumptions, it is natural to consider model \eqref{haf} near QCP using methods which were successfully applied before for discussion of diluted HAFs near percolation threshold at zero magnetic field (see Refs.~\cite{percrev,harkir,shenderaf,shenderfm} and references therein). We remind first briefly the main aspects of that analysis and point out then similarities and distinctions between diluted HAF and our system. 

Previous considerations \cite{percrev,harkir,shenderaf,shenderfm} of diluted HAFs rely heavily on the assumption that low-energy elementary excitations in the infinite network near percolation threshold are weakly damped gapless spin waves (hydrodynamic excitations). The existence of the hydrodynamic excitations is closely related to commutativity of the Hamiltonian with the total spin operator. To the best of our knowledge, $d$ value has not been established above which the hydrodynamic description of excitations is correct in diluted HAFs. It is well known that it is valid in diluted 3D HAFs up to the percolation threshold. \cite{percrev} On the other hand, it is found in Ref.~\cite{2dvac} that a small concentration $c$ of vacancies in 2D HAFs leads to a localization of spin waves with wavelengths greater than $e^{\pi/4c}$ (see also Ref.~\cite{harkir}). 

We rely below on the assumption that well-defined long-wavelength quasiparticles exist at $h\alt h_c$. Bearing in mind results of Ref.~\cite{2dvac} and a good agreement of formulas obtained below to numerical findings at $d=3$, we indicate $d>2$ as a rough estimation of the range of our results validity. Indeed, further consideration is required for a more accurate determination of the $d$ value above which our discussion is valid.

\subsection{Diluted HAF near percolation threshold}
\label{dilhaf}

The hydrodynamic excitations can be described phenomenologically using the following expression for the system energy $E$ in the continuum limit accounting for fluctuations in transverse components of sublattices magnetizations ${\bf m}_1({\bf r})$ and ${\bf m}_2({\bf r})$: \cite{harkir}
\begin{equation}
\label{e}
E = \int d{\bf r} 
\sum_\alpha
\left(
\frac{\Upsilon}{4M^2} 
\left(
\overrightarrow{\nabla} m_1^\alpha({\bf r}) - \overrightarrow{\nabla} m_2^\alpha({\bf r})
\right)^2
+
\frac{H_e}{2M} \left( m_1^\alpha({\bf r}) + m_2^\alpha({\bf r})\right)^2
\right),
\end{equation}
where $\alpha=$ \texttt{x,y} at zero magnetic field, \texttt{z} axis is directed along sublattices magnetizations, we consider 3D HAF for definiteness, and $\Upsilon$, $M$, and $H_e$ are phenomenological constants which have to be found from an analysis of the corresponding microscopic theory. 

$M$ is a mean staggered magnetization per unit volume which is proportional to the number of sites in the infinite network. \cite{harkir} $M$ scales near the percolation threshold $x_c$ as
\begin{equation}
\label{m}
M\propto (x-x_c)^{\beta}.
\end{equation}

$\Upsilon$ is a measure of the energy needed to create a spatial variation of the staggered magnetization (i.e., $\Upsilon$ is the helicity modulus). It is related to the conductivity of the system which is obtained from the diluted HAF by replacing bond between neighboring sites $i$ and $j$ by a resistor with conductance $\sigma_{ij}=J_{ij}$ (in diluted HAF, $J_{ij}$ is taken to be $Jp_ip_j$, where $p_i=1$ for an occupied site and $p_i=0$ for a vacancy): \cite{harkir}
\begin{equation}
\label{ups}
\Upsilon \propto (x-x_c)^{\sigma}.	
\end{equation}
To demonstrate this, periodic boundary conditions are applied along \texttt{x} and \texttt{z} directions and open boundaries are assumed in the \texttt{y} direction. Let us consider a spin fluctuation having the form
\begin{equation}
\label{fluc}
{\bf S}_j = {\bf M}_j \left(1-\theta_j^2/2\right) + e^{i{\bf k}_0{\bf R}_j}\theta_j {\bf n} ,
\end{equation}
where ${\bf k}_0=(\pi,\pi,\pi)$ is the antiferromagnetic vector, $e^{i{\bf k}_0{\bf R}_j}$ is equal to $+1$ and $-1$ on sites from different sublattices, ${\bf M}_j=e^{i{\bf k}_0{\bf R}_j}\hat z$, $\hat z$ is the unit vector directed along \texttt{z} axis, and $\bf n$ is a unit vector from \texttt{xy} plane. Fixing $\theta_i=\theta_0$ on one open boundary and $\theta_i=-\theta_0$ on another boundary, we obtain for the equilibrium conditions at site $i$ lying inside the system
\begin{equation}
\label{kir}
\sum_j J_{ij}(\theta_i-\theta_j)=\sum_j \sigma_{ij}(\theta_i-\theta_j)=0.
\end{equation}
Eqs.~\eqref{kir} together with the boundary conditions are equivalent to Kirchoff's law of the network of conductances $\sigma_{ij}$ connecting electrodes of potentials $\theta_0$ and $-\theta_0$. \cite{harkir} 

Applying a small uniform transverse field and minimizing the energy $E$, one finds that $H_e=M/\chi_\perp$, where $\chi_\perp$ is a transverse susceptibility. It is demonstrated in Ref.~\cite{harkir} that $\chi_\perp \propto (x-x_c)^{-\gamma}$ so that
\begin{equation}
\label{he}
H_e \propto (x-x_c)^{\gamma+\beta}.	
\end{equation}
Landau-Lifshitz equations
\footnote{The effective fields acting on ${\bf m}_{1,2}({\bf k})$ which appear in Landau-Lifshitz equations are determined as $\frac{(2\pi)^d}{V} \partial E/\partial {\bf m}_{1,2}({\bf k})$, where $V$ is the system volume.} 
for Fourier components of ${\bf m}_1({\bf r})$ and ${\bf m}_2({\bf r})$ give the spectrum of the doubly degenerate (due to fluctuations along equivalent \texttt{x} and \texttt{y} directions) Goldstone mode \cite{harkir}
\begin{equation}
\label{specaf}
\begin{aligned}
	\epsilon_{\bf k} &= Ck,\\
	C &= \sqrt{2\Upsilon/\chi_\perp} \propto (x-x_c)^{(\sigma+\gamma)/2}.	
\end{aligned}
\end{equation}
These propagating excitations exist on the length scale greater than the correlation length $\xi\propto|x-x_c|^{-\nu}$. On smaller length scale, excitations (the so-called ``fractons'') are localized. \cite{percrev}

\subsection{Model \eqref{haf} near QCP}
\label{genqcp}

\begin{figure}
\includegraphics[scale=0.8]{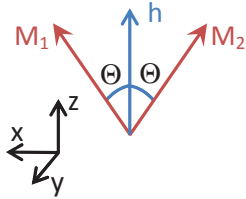}
\caption{(Color online). Sublattices magnetizations (per unit volume) ${\bf M}_{1,2}$ in external magnetic field $\bf h$.
\label{axes}}
\end{figure}

As the infinite network in our system is surrounded by areas having saturated magnetization and a gapped spectrum (the gap in such areas depends linearly on $h$), the system resembles the diluted HAF in many respects. In particular, we can analyze the low-energy dynamics as it is done above which is governed by low-energy excitations in the infinite network and large finite clusters. The difference arises from the fact that HAF sublattices are not collinear and they are directed by an angle $\Theta$ to the magnetic field (see Fig.~\ref{axes}). $\Theta$ is an extra phenomenological parameter in the theory. As soon as the infinite network breaks up into finite clusters which magnetizations are not saturated, one concludes that $\Theta$ is finite at $h\le h_c$. 

It can be easily shown that the energy of fluctuations in the transverse components of the sublattices magnetizations is given by Eq.~\eqref{e} as in diluted HAF (the angle $\Theta$ arises only in Landau-Lifshitz equations). However fluctuations within \texttt{xz} plane (within which the field and sublattices magnetizations lie) and those along \texttt{y} direction have different properties. Analysis of the Landau-Lifshitz equations shows that the former have gapped spectrum whereas the latter are characterized by the Goldstone mode with dispersion \eqref{specaf}, where $C$ should be multiplied by $\sin\Theta$. We imply below that $\alpha=$ \texttt{y} in Eq.~\eqref{e} in the Goldstone mode discussion. 

Phenomenological constants $M$ and $\Upsilon$ have the same meaning as in diluted HAFs and we try them in the form \eqref{m} and \eqref{ups}, respectively. Due to the finiteness of $\Theta$ at $h=h_c$, $\beta$ is the critical exponent of the order parameter. The equilibrium condition for fluctuations of the type \eqref{fluc} (where $\bf n$ is directed along \texttt{y} axis, $\theta_j$ is the angle between $j$-th magnetic moment and \texttt{xz} plane, and ${\bf M}_j=(e^{i{\bf k}_0{\bf R}_j}\sin\Theta,0,\cos\Theta)$) reads as (cf.\ Eq.~\eqref{kir})
\begin{equation}
\label{kir2}
	\sum_j 
	J_{ij} \left(\theta_i-\theta_j - \theta_i \cos\phi_i \frac{\sin(\phi_i-\phi_j)}{\sin\phi_i}\right)
	= 0,
\end{equation}
where $\phi_i$ is the angle between $i$-th magnetic moment and the field $\bf h$ and $i$ is a site from the infinite network. The last term is negligible in Eq.~\eqref{kir2} and the analog of the Kirchoff's law is recovered provided that $|\phi_i-\phi_j|\ll|\sin\phi_i|$. The latter inequality is expected to hold at least not very close to the infinite network edges. Then, it seems to us likely that the helicity modulus is related to the conductivity of the corresponding system however further consideration may be required to proof it rigorously. We use below the relation between $\Upsilon$ and the conductivity only in the discussion of the possible percolation regime.

$H_e$ is related in our system to the susceptibility $\chi_y$ to a small field directed along \texttt{y} axis. The important difference from diluted HAFs is that $\chi_y$ remains finite at $h=h_c$. Indeed, the origin of $\chi_\perp$ divergence in diluted HAFs is uncompensated net spins arising on large length scales which lead to large magnetization of the whole system in the external field (see Refs.~\cite{harkir,percrev} for more details). In contrast, there are no such objects in our system and a small additional field along \texttt{y} axis leads only to a small turn of the external field in the \texttt{y} direction. This rotation produces just a small magnetization along \texttt{y} direction proportional to the magnetization of the whole sample. As a result, one arrives at Eq.~\eqref{he} for $H_e$, where 
\begin{equation}
\gamma=0.
\end{equation}
This circumstance leads to some important differences in dynamical properties of our system in comparison with diluted HAFs.

DOS of the infinite network can be found as it was done in Refs.~\cite{harkir,shenderaf,percrev} for diluted HAFs, the result being
\begin{eqnarray}
\label{dosaf}
{\cal D}_{inf}(\omega) 
&\propto &
\left\{
\begin{aligned}
&(h_c-h)^{\beta}\omega^{v_1}, \quad && \omega\gg\omega_0\sim(h_c-h)^{\nu+\sigma/2},	\\
&\frac{\omega^{d-1}}{(h_c-h)^{d\sigma/2}}, \quad && \omega\ll\omega_0,	
\end{aligned}
\right.\\
\label{v1}
v_1 &=& 2\frac{d\nu-\beta}{\sigma+2\nu}-1.
\end{eqnarray}
One can further simplify Eq.~\eqref{v1} using the geometrical relation $D\nu=d\nu-\beta$, \cite{foot} where $D<d$ is the fractal dimension (i.e., the space dimension on the length scale smaller than $\xi$). The second line in Eq.~\eqref{dosaf} is simply DOS of propagating excitations \eqref{specaf}. The quantity $\omega_0$ is given by Eq.~\eqref{specaf} at $k\sim1/\xi$. The first line in Eq.~\eqref{dosaf} is found by taking into account that properties of excitations (energy and DOS normalized to one spin) on the length scale smaller than $\xi$ (corresponding to $\omega\gg\omega_0$) do not depend on the proximity to QCP (i.e., on $h_c-h$). \cite{shenderaf,halph} Then, DOS of localized excitations can be represented as $(h_c-h)^{\beta}\omega^{v_1}$ that should match DOS of the hydrodynamic mode at $\omega\sim\omega_0$. As a result one arrives at Eq.~\eqref{v1} for $v_1$.

The dynamical critical exponent $z$ is determined by the scaling of $\omega_0$ (see Eq.~\eqref{dosaf}) that gives 
\begin{equation}
\label{z}
	z = 1 + \frac{\sigma}{2\nu}.
\end{equation}
Because $\gamma=0$ in our system, Eqs.~\eqref{dosaf}--\eqref{z} differ from their counterparts in diluted HAFs.

To estimate the N\'eel temperature, we follow Ref.~\cite{shenderaf} and introduce first an auxiliary quantity
\begin{equation}
\label{phidef}
\Phi(k) = \frac{E(k)}{\left|{\bf m}_1\left({\bf k}\right)\right|^2},
\end{equation}
where $E(k)$ is the energy \eqref{e} of the spin density wave characterized by momentum $\bf k$. To find it at $k<1/\xi$, we express ${\bf m}_2({\bf k})$ via ${\bf m}_1({\bf k})$ using Landau-Lifshitz equations, substitute the result to Eq.~\eqref{e} and obtain $\Phi(k) = \Upsilon k^2/M^2$ (the derived expression differs from that in diluted HAF \cite{shenderaf} by a factor of 2 due to different number of Goldstone modes). As soon as $k(\omega)=\omega/C$, one easily finds from this result $\Phi(\omega)$ at $\omega\ll\omega_0$. As soon as $\Phi(\omega)$ does not depend on $h_c-h$ at $\omega\gg\omega_0$, we try $\Phi(\omega)$ in the form $\omega^{v_\phi}$ in this regime, where $v_\phi$ is found by matching $\Phi(\omega)$ at $\omega\sim\omega_0$. One obtains as a result
\begin{eqnarray}
\label{phi}
\Phi(\omega) 
&\propto &
\left\{
\begin{aligned}
& \omega^{v_\phi}, \quad && \omega\gg\omega_0,	\\
& \omega^2 (h_c-h)^{-2\beta}, \quad && \omega\ll\omega_0,	
\end{aligned}
\right.\\
v_\phi &=& 2-\frac{4\beta}{\sigma+2\nu}.
\end{eqnarray}
The reduction of sublattices magnetization $\delta M$ due to thermal fluctuations at small temperature $T\ll T_N(h)$ at a given $h<h_c$ is estimated as \cite{shenderaf}
$
\delta M/M\sim \langle \sum_{\bf k}|{\bf m}_1({\bf k})|^2\rangle/2M^2
=
\langle \sum_{\bf k}E(k)/\Phi(k)\rangle/2M^2
$,
where we use Eq.~\eqref{phidef} and $\langle\dots\rangle$ denotes thermal average. Replacing the summation over momenta by integration over energy, one obtains \cite{shenderaf}
\begin{equation}
\label{dm}
\frac{\delta M}{M} = \frac{1}{2M^2} 
\int\frac{\omega{\cal D}_{inf}(\omega)}{(e^{\omega/T}-1)\Phi(\omega)}d\omega.
\end{equation}
We find from Eqs.~\eqref{dosaf}, \eqref{phi}, and \eqref{dm} that $\delta M\ll M$ at $T\ll\omega_0$. At larger temperatures, $\omega_0\ll T\ll T_N(h)$, the region of $\omega\sim\omega_0$ gives the main contribution to the integral in Eq.~\eqref{dm} provided that 
\begin{equation}
\label{betain}
	\beta<\sigma+(2-d)\nu. 
\end{equation}
We assume below that this inequality is fulfilled while it is not generally the case. For instance, Eq.~\eqref{betain} holds in the percolation theory at $d=2,3$ but it does not hold at $d\ge6$ (i.e., for the mean-field critical percolation exponents). \cite{aha} As will be seen below, Eq.~\eqref{betain} is fulfilled in the close vicinity of QCP at $d>2$. As a result, one finds from Eq.~\eqref{dm} $\delta M/M\sim T (h_c-h)^{(d-2)\nu-\sigma}$. As soon as this expression is valid also by the order of magnitude at $T\sim T_N(h)$, when $\delta M\sim M$, we obtain the following estimation for the N\'eel temperature:
\begin{equation}
\label{tn}
\begin{aligned}
T_N(h) &\propto (h_c-h)^{\phi},\\
\phi &= \sigma+(2-d)\nu.
\end{aligned}
\end{equation}
The corresponding expression is also readily written out when Eq.~\eqref{betain} does not hold.

The order parameter static correlation function having the form $\chi(r)\sim r^{-(d-2+z+\eta)}$ at QCP (see Ref.~\cite{Fisher}) should behave as the fractal correlation function $g(r)\sim r^{-2(d-D)}$ giving the probability that a site a distance $r$ apart from the given site belongs to the same cluster. \cite{foot} Using also the geometrical relation $D\nu=d\nu-\beta$ one derives
\begin{equation}
\label{eta}
\eta = \frac{2\beta}{\nu}+2-d-z.
\end{equation}
Eq.~\eqref{eta} is obtained in Ref.~\cite{Fisher} using scaling relations and implying that $\beta$ is the order parameter exponent.

One needs also DOS of finite clusters ${\cal D}_{fin}(\omega)$ to describe the system behavior in the ordered phase near QCP. According to our assumptions, one can find finite clusters of all characteristic sizes smaller than $\xi$ near QCP while the probability to find larger clusters is exponentially small. As a result, ${\cal D}_{fin}(\omega)=0$ at $\omega\ll\omega_0$. Then, one try ${\cal D}_{fin}(\omega)$ in the form $\omega^{v_2}$ at $\omega\gg\omega_0$ because the number of sites in finite clusters does not vanish at $h=h_c$. To find $v_2$, one notices that DOS per spin is of the order of $\omega_0^{v_1}$ at $\omega\sim\omega_0$ in the infinite network (see Eq.~\eqref{dosaf}). DOS per spin should have the same form in the largest clusters which characteristic linear size is of the order of $\xi$. Thus, one arrives at the estimation 
$
V_\xi\omega_0^{v_1} \sim {\cal D}_{fin}(\omega\sim\omega_0) \propto \omega_0^{v_2}
$, 
where $V_\xi$ is the total volume of the largest clusters. Because the infinite network breaks predominantly into finite clusters with linear size of the order of $\xi$, $V_\xi$ is of the order of the infinite network volume $(h_c-h)^\beta$. One obtains as a result
\begin{eqnarray}
\label{dosfin}
{\cal D}_{fin}(\omega) 
&\propto &
\left\{
\begin{aligned}
&\omega^{v_2}, \quad &&\omega\gg\omega_0,	\\
&0, \quad &&\omega\ll\omega_0,
\end{aligned}
\right.\\
\label{v2}
v_2 &=& \frac{d-z}{z},
\end{eqnarray}
where Eq.~\eqref{z} is also taken into account. 

\section{Close vicinity of QCP}
\label{qcp}

In the dirty-boson problem considered in Ref.~\cite{Fisher}, the equality
\begin{equation}
\label{z=d}
	z=d
\end{equation}
comes from the finiteness of the compressibility at QCP. Arguments against that derivation are presented in Ref.~\cite{weich}. In Ref.~\cite{yu2}, Eq.~\eqref{z=d} is derived in a different way from the scaling of the particle-hole gap. Our consideration above suggests another way to obtain this equality. The counterpart of the compressibility in our system is the longitudinal susceptibility $\chi_\|(h)=-d^2E(h)/dh^2$. It is clear from a physical point of view that $\chi_\|(h_c)$ is a constant because a large amount of arbitrary large clusters with unsaturated magnetization remains in the system. But we cannot relate $\chi_\|(h)$ to any quantity consider above using just its definition because we discuss only energy change by fluctuations in the direction transverse to the field. However, one can obtain Eq.~\eqref{z=d} by assuming that the singular part of the free energy $f_s$ scales inversely with the space-time correlation volume, i.e., $f_s\sim(h_c-h)^{\nu(d+z)}$. \cite{weich,yu2,Fisher} This assumption has not been questioned yet in close vicinity of QCP. One has as a result for the helicity modulus \cite{Fisher} $\Upsilon\propto (h_c-h)^{\nu(d+z-2)}$ so that $\sigma=\nu(d+z-2)$ (see Eq.~\eqref{ups}). \cite{foot2} Substituting the latter equality to Eq.~\eqref{z} we come to Eq.~\eqref{z=d}. 

One obtains from Eqs.~\eqref{z}, \eqref{tn}, and \eqref{z=d}
\begin{eqnarray}
\label{sigma2}
\sigma &=& 2\nu(d-1),\\
\label{phi2}
\phi &=& \nu d.
\end{eqnarray}

It is seen from Eqs.~\eqref{dosfin}--\eqref{z=d} that DOS of finite clusters is a constant at $\omega\gg\omega_0$. In particular, ${\cal D}_{fin}(\omega)=\rm const$ for $\omega\ge0$ at QCP. As a consequence, one derives for the specific heat $\cal C$ at $h=h_c$
\begin{eqnarray}
\label{c}
{\cal C}  &\propto &
\frac{d}{dT}\int\frac{\omega{\cal D}_{fin}(\omega)}{e^{\omega/T}-1}d\omega
\sim T.
\end{eqnarray}

One recognizes in Eqs.~\eqref{sigma2} and \eqref{phi2} expressions for critical exponents of the helicity modulus and the critical temperature obtained using the scaling ansatz in Ref.~\cite{Fisher}. The relation ${\cal C}\propto T$ is also found in Ref.~\cite{Fisher} using scaling arguments. Then, we come to the same results from another direction. We are able also to derive some new expressions. It is argued in Ref.~\cite{Fisher} that one-particle DOS scales as $\omega^{(d-2+\eta)/z}$ at QCP so that our finding ${\cal D}_{fin}(\omega)=\rm const$ signifies that
\footnote{Remember that DOS in disordered HAFs corresponds to DOS of uncondensed particles in a system of dirty bosons. However, the condensate disappears at QCP so that this comparison of DOS is correct.}
\begin{equation}
\label{eta2}
	\eta=2-d
\end{equation}
as in disordered Fermi systems (see Ref.~\cite{Fisher} for a more detailed discussion of the similarity between disordered Fermi and Bose systems). Notice that inequality $2-2d<\eta\le 2-d$ is proposed in Ref.~\cite{Fisher} that is valid, e.g., at $d=1$ in which case $\eta=1/3$. \cite{shulz} Then, we establish that $\eta$ coincides with the upper bound of this interval at $d>2$. One obtains from Eqs.~\eqref{eta} and \eqref{eta2}
\begin{equation}
\label{beta}
\beta = \frac{\nu d}{2}.
\end{equation}
To the best of our knowledge, Eqs.~\eqref{eta2} and \eqref{beta} have not been obtained before. Notice that Eqs.~\eqref{phi2} and \eqref{beta} are consistent with values of $\phi=2.7(2)$, $\nu=0.88(5)$, and $\beta=1.5(2)$ found in Ref.~\cite{kisel} for $d=3$. Table~\ref{table2} summarizes all expressions for critical exponents derived in this section and demonstrates the consistency of the theory at $d=3$ with numerical results of Ref.~\cite{kisel}.

\begin{table}
\caption{Expressions for critical exponents obtained in the present consideration at $d>2$ in close vicinity of QCP (expressions for $\phi$, $z$, and $\sigma$ coincide with those obtained in Ref.~\cite{Fisher}). Predictions of the theory for $d=3$ are compared with numerical results of Ref.~\cite{kisel}. Values in the column "theory" are found by substituting $\nu=0.88(5)$ (observed in Ref.~\cite{kisel}) to the expressions.
\label{table2}
}
\begin{ruledtabular}
\begin{tabular}{|c|ccc|}
				& & numerical values at $d=3$ &  \\
        & theory & & numeric \\
\hline
$\nu$ 					& --- & & $0.88(5)$  \\
$\phi=\nu d$ 		& $2.6(2)$ & & $2.7(2)$  \\
$z=d$ 					& 3 & & 3 \\
$\sigma=2\nu(d-1)$ 	& 3.5(2) & & --- \\
$\beta=\nu d/2$ & $1.32(8)$ & & 1.5(2) \\
$\eta=2-d$			& $-1$ & & ---
\end{tabular}
\end{ruledtabular}
\end{table}

It should be pointed out that Eqs.~\eqref{z=d}, \eqref{sigma2}, and \eqref{beta} simplify greatly some general relations obtained in Sec.~\ref{genqcp}. In particular, one finds 
\begin{equation}
\label{vv}
	v_1=-1/2 \ \mbox{  and  }\ v_2=0
\end{equation}
which appear to be independent of $d$. Thus, similar to fractons in diluted HAFs at $d\ge2$ (see Ref.~\cite{percrev}), DOS per spin of fractons both in the infinite network and in finite clusters shows in our system {\it superuniversal} properties at $d>2$.

\section{Possible crossover to percolation scenario near QCP at $d=3$}
\label{seccross}

It should be noted that the results obtained in Ref.~\cite{kisel} which are consistent with formulas discussed in the previous section are obtained in 3D models, where the disorder strength $\Delta$ is a driving parameter, the average chemical potential $\mu$ is fixed, and the transition takes place at quite large $\Delta$. On the other hand there is a large amount of results obtained experimentally and numerically in another part of the phase diagram in the plane $\Delta$--$\mu$ (at smaller $\Delta$) by varying $\mu$ at fixed $\Delta$ (this part of the phase diagram is also examined in Ref.~\cite{kisel} on larger systems, main findings being consistent with those from previous works). \cite{critkis} These results are in sharp contradiction with Ref.~\cite{Fisher}. We show now that they could be described by formulas obtained in Sec.~\ref{genqcp} using percolation critical exponents. This points to a possibility of a crossover in the critical region in a part of the phase diagram from the percolation regime to that predicted in Ref.~\cite{Fisher} and discussed in the previous section.

Notice first that values of the critical exponent $\nu$ of the correlation length obtained in all previous numerical works \cite{Yu,yu2,haas,hitch,kisel} are close to the percolation one $\nu_p=0.876$. On the other hand, the value of $\nu=0.88(5)$ obtained in Ref.~\cite{kisel} very close to QCP is equal to within the error to $\nu_p$.  Then, the necessary (but not sufficient) condition seems to be fulfilled for the possibility of the crossover scenario in the critical region.

\begin{table}
\caption{Comparison of critical exponents obtained before experimentally and numerically in various 3D systems with corresponding values in the percolation regime discussed in Sec.~\ref{seccross} (percolation exponents have index $p$). Values of percolation critical exponents are taken from Refs.~\cite{percexp,percexp2,percexp3}.
\label{table}
}
\begin{ruledtabular}
\begin{tabular}{|c|c|ccc|}
critical & expression via percolation & & numerical values &  \\
exponent & critical exponents        & percolation & experiment & numerics \\
\hline
$\nu$ & $\nu_p$ & 0.876 & not available & $0.70(15)$ (Refs.~\cite{hitch,yu2,haas}) \\
$\beta$ & $\beta_p$ & 0.418 & $0.4\div 0.5$ (Refs.~\cite{Huv1,wulf}) & 
0.6(1) (Ref.~\cite{kisel}) \\
$\phi$ &  $\sigma_p-\nu_p$ & $1.124$ & $1.1\div1.2$ (Refs.~\cite{Yu,Huv1,yama}) & $1.1\div1.2$ (Refs.~\cite{Yu,yu2,haas,critkis,kisel}) \\
$z$ & $1+\sigma_p/2\nu_p$ & $2.142$ & not available & $\approx 3$ (Refs.~\cite{hitch,yu2}); $1.4\div3^{a)}$
\end{tabular}
\end{ruledtabular}
$^{a)}$ Numerical data from Ref.~\cite{yu2} are consistent also with these values (see Appendix~\ref{mc}).
\end{table}

Critical exponents calculated from Eqs.~\eqref{z} and $\eqref{tn}$ using percolation critical exponents (assuming that $\sigma$ is the critical exponent of the conductivity) are compared in Table~\ref{table} with results of previous experimental and numerical works. It is seen that all experimental and many numerical findings are consistent with the percolation scenario. However some remarks are in order. 

First, some previous numerical calculations \cite{yu2,haas} give values of $\beta\approx 1$ which are substantially greater than $\beta_p\approx0.42$. However consideration of the corresponding model in Ref.~\cite{kisel} gives the value of $\beta=0.6(1)$ not very close to QCP which is much closer to $\beta_p$. Notice that system sizes in Ref.~\cite{kisel} are at least an order of magnitude larger than in previous numerical works. Then, there is a tendency for $\beta$ decreasing upon the system size increasing. That is why we indicate the value of $0.6(1)$ for $\beta$ in Table~\ref{table}. 

Second, values of $z\approx3$ were reported before for 3D systems \cite{yu2,haas,hitch} which are larger than the proposed percolation value $1+\sigma_p/2\nu_p\approx 2.14$. However we demonstrate in Appendix~\ref{mc} that numerical data taken from Ref.~\cite{yu2} are not very sensitive to a variation in $z$ in quite a broad interval (this interval is $(1.4,3)$ for some data). It should be stressed also that equality \eqref{z=d} holds only in the close vicinity of QCP so that it can be violated in the percolation regime. Certainly, we are unable to reexamine all relevant data from all previous numerical papers. We choose Ref.~\cite{yu2} for this purpose recognizing that it is one of the most detailed and reliable works. Further numerical consideration may be necessary to find $z$ value not very close to QCP.

As it is shown above, the common assumption that the singular part of the free energy scales inversely with the space-time correlation volume $f_s\sim(h_c-h)^{\nu(d+z)}$ leads to Eq.~\eqref{z=d} that would be violated in the percolation regime. To be consistent with Eq.~\eqref{z}, $f_s$ would scale as 
\begin{equation}
\label{fs}
	f_s \sim (h_c-h)^{2z\nu}
\end{equation}
not very close to QCP. In accordance with Eq.~\eqref{z}, Eq.~\eqref{fs} signifies that $\sigma=2\nu(z-1)$. \cite{foot2} Notice that Eq.~\eqref{fs} works also close to QCP, where $z=d$. The finite-size scaling would change accordingly for the helicity modulus:
\begin{equation}
\Upsilon(h,L) = L^{-(2z-2)} F((h_c-h)^\nu L).
\end{equation}

DOS per spin would not be superuniversal in the percolation regime: $v_1$ and $v_2$ given by Eqs.~\eqref{v1} and \eqref{v2} do depend on $d$.

\section{Summary and conclusion}
\label{conc}

In summary, we discuss the magnetically ordered phase of HAF \eqref{haf} with bond disorder near quantum phase transition to the disordered BG phase. Our consideration is based on the assumption that long-wavelength quasiparticles are well-defined at $h\le h_c$. Besides, we assume that the transition bear a qualitative resemblance to the percolation one. We derive formulas for the N\'eel temperature \eqref{tn}, the dynamical critical exponent \eqref{z}, DOS of infinite network \eqref{dosaf} and DOS of finite clusters \eqref{dosfin} which are valid in the entire critical region. In the close vicinity of QCP, where Eq.~\eqref{z=d} holds, we recover by these formulas some results obtained before using physical arguments and the scaling ansatz suggested in Ref.~\cite{Fisher} (see Eqs.~\eqref{eta}, \eqref{sigma2}--\eqref{c}). As an extension of the previous analysis, we derive Eqs.~\eqref{eta2} and \eqref{beta} (the latter is in quantitative agreement with Ref.~\cite{kisel}). DOS per spin of localized excitations (fractons) are found to be superuniversal (i.e., independent of $d$) both in the infinite network and in finite clusters (see Eq.~\eqref{vv} and first lines in Eqs.~\eqref{dosaf} and \eqref{dosfin}). An alternative way is proposed to derive the equality $z=d$ close to QCP. We show that all experimental data and many numerical ones obtained before in various 3D systems would be relevant to the percolation critical scenario which could arise not very close to QCP before the onset of the regime predicted in Ref.~\cite{Fisher}. However further numerical and experimental considerations are necessary to support this idea. The results obtained are expected to be valid starting from a $d$ value which lies between 2 and 3 and determination of which is a subject of future work.

\begin{acknowledgments}

This work is supported by Russian Science Foundation (grant No.\ 14-22-00281).

\end{acknowledgments}

\appendix

\section{Dynamical critical exponent in previous numerical consideration}
\label{mc}

In this appendix, we revisit numerical data obtained in Ref.~\cite{yu2} by quantum Monte Carlo (QMC) simulations. The system under consideration is widely used model of the DTN compound ({NiCl$_{2} \cdot$ 4SC(NH$_2$)$_2$}). DTN is an antiferromagnet whose Hamiltonian has the form \eqref{haf} in which a large (compared to the exchange coupling constants) single-ion easy-plane anisotropy is added having the form $D \sum_i(S^{z}_i)^{2}$, where $D>0$. At low temperature, pure DTN has two QCPs at fields $h_{c1}$ and $h_{c2}>h_{c1}$. A magnetically ordered (``SF'') phase with canted antiferromagnetic order arises at $h_{c1}<h<h_{c2}$ whereas ``MI'' (paramagnetic) phases are stable outside this interval. Disorder leads to BG phases near both critical fields. Both transitions from BG to ``SF'' phases are discussed in Ref.~\cite{yu2}, where two types of quenched disorder are considered: 1) random substitution of some Cl atoms by Br ones (the so-called Br-DTN) that changes exchange constants and $D$ values around the defect; 2) substitution of magnetic Ni ions by non-magnetic ones (i.e., dilution of the magnetic subsystem). We revisit QMC data only for the case of dilution because the finite-size scaling in Br-DTN is somewhat unsatisfactory as it is explained in Ref.~\cite{yu2}. 

\begin{figure}
\includegraphics[scale=0.65]{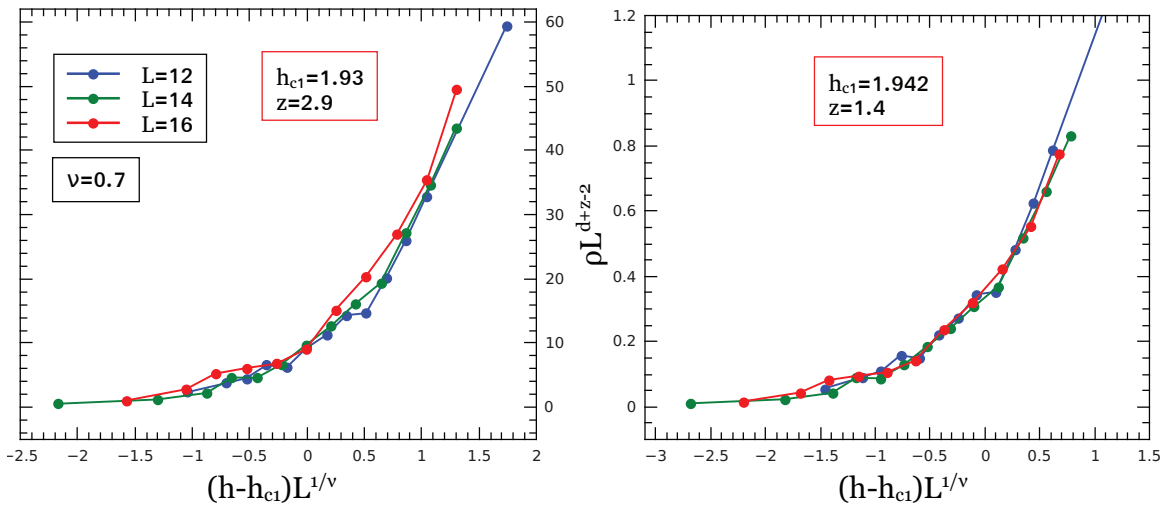}
\caption{(Color online). Finite-size scaling plots for the spin stiffness $\rho$ in the site-diluted DTN at $h\approx h_{c1}$. Numerical data are taken from Fig.~8 of Ref.~\cite{yu2}. Left panel is drawn with parameters from original paper \cite{yu2} ($h_{c1}=1.93$, $\nu=0.7$, and $z=2.9$). We change values of $h_{c1}$ and $z$ in the right panel in comparison with the left one.
\label{fig1}}
\end{figure}

\begin{figure}
\includegraphics[scale=0.65]{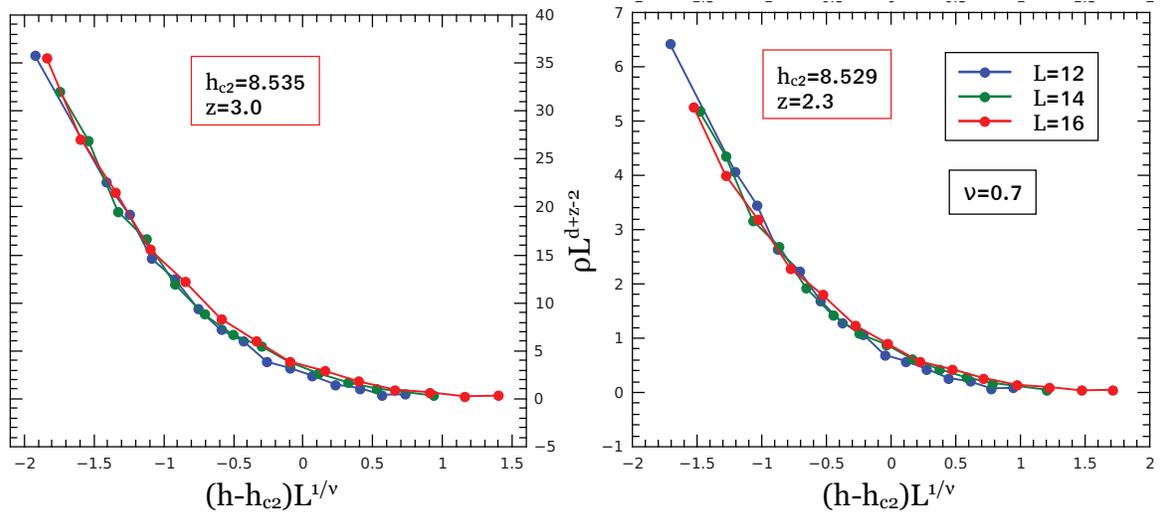}
\caption{(Color online). Same as in Fig.~\ref{fig1} but for $h\approx h_{c2}$.
\label{fig2}}
\end{figure}

The dynamical critical exponent $z$ is extracted in Ref.~\cite{yu2} from finite-size scaling for the spin stiffness (the helicity modulus) $\rho$. Corresponding plots presented in Fig.~8 of Ref.~\cite{yu2} for the site-diluted DTN with dilution $x=0.15$ are shown in left panels of Figs.~\ref{fig1} and \ref{fig2}. In right panels of those figures, we present the same data drawn using smaller $z$ values. We also shift a little values of $h_{c2}$ and $h_{c1}$ in right panels so that they remain consistent with those found in Ref.~\cite{yu2} in corresponding analysis of data for the correlation length and the order parameter. It is seen that collapses of data obtained for different system sizes $L^3$ in right panels are at least no worse than those in left panels. This signifies that smaller $z$ values are also consistent with numerical data.

\bibliography{BGSLbib}

\begin{thebibliography}{30}
\bibitem[{foo({\natexlab{a}})}]{foot}
\bibinfo{howpublished}{One can obtain this equality as it is done in
  percolation theory. \cite{aha,percrev} Notice that this derivation is valid
  in our system too under assumptions discussed in Sec.~\ref{prel}.}
\bibitem[{foo({\natexlab{b}})}]{foot2}
\bibinfo{howpublished}{Imposition of a gradient $\nabla\varphi$ of the order
  parameter phase along one of the spatial directions leads to a free energy
  difference $\Delta f_s\propto \Upsilon (\nabla\varphi)^2$. Since
  $\nabla\varphi$ has dimension of inverse length implying
  $\nabla\varphi\sim1/\xi$, it follows that $\Upsilon\propto
  (h_c-h)^{\nu(d+z-2)}$. \cite{Fisher}}.
\end{thebibliography}

\end{document}